\documentclass[useAMS,usenatbib]{mn2e}
\usepackage{graphicx}
\usepackage{epstopdf}
\newcommand{\araa}{Annual Review of Astron and Astrophys}

\newcommand{\aap} {Astronomy \& Astrophysics}

\newcommand{\aj}{Astronomical Journal}
\newcommand{\apj}{Astrophysical Journal}
\newcommand{\apjl}{Astrophysical Journal, Letters}
\newcommand{\apjs}{Astrophysical Journal, Supplement}
\newcommand{\mnras}{Monthly Notices of the RAS}

\title[GAMA: Low H$\alpha$ luminosity galaxies]{Galaxy And Mass Assembly (GAMA): Galaxies at the faint end of the H$\alpha$ luminosity function}
\author[S. Brough et al.]{S. Brough,$^{1}$\thanks{E-mail: sb@aao.gov.au}, 
A.M.~Hopkins$^1$,
R.G.~Sharp$^1$,
M.~Gunawardhana$^{1,2,3}$,
D.~Wijesinghe$^{3}$,
\newauthor A.S.G.~Robotham$^4$,
S.P.~Driver$^4$,
I.K.~Baldry$^5$,
S.P.~Bamford$^6$,
J.~Liske$^7$,
J.~Loveday$^8$,
\newauthor P.~Norberg$^9$,
J.A.~Peacock$^9$,
J.H.~Bland-Hawthorn$^{3}$,
M.J.I.~Brown$^{10}$,
E.~Cameron$^{11}$,
\newauthor S.M.~Croom$^{3}$,
C.S.~Frenk$^{12}$,
C. Foster$^{13}$,
D.T.~Hill$^4$,
D.H.~Jones$^1$,
L.S.~Kelvin$^4$,
\newauthor  K.~Kuijken$^{14}$,
R.C.~Nichol$^{15}$,
H.R.~Parkinson$^9$,
K. Pimbblet$^{10}$,
C.C.~Popescu$^{16}$,
\newauthor  M.~Prescott$^5$,
W.J.~Sutherland$^{17}$,
E.~Taylor$^{3}$,
D.~Thomas$^{15}$,
R.J.~Tuffs$^{18}$,
\newauthor E.~van~Kampen$^7$ \\
$^1$Australian Astronomical Observatory, PO Box 296, Epping, NSW 1710, Australia\\
$^{2}$Department of Physics \& Astronomy, Macquarie University, NSW  2109, Australia\\
$^{3}$Sydney Institute for Astronomy, School of Physics, University of Sydney, NSW 2006, Australia\\
$^4$School of Physics \& Astronomy, University of St Andrews, North Haugh, St Andrews, KY16 9SS, UK, SUPA\thanks{Scottish Universities Physics Alliances}\\
$^5$Astrophysics Research Institute, Liverpool John Moores University, Twelve Quays House, Egerton Wharf, Birkenhead, CH41 1LD, UK\\
$^6$Centre for Astronomy and Particle Theory, University of Nottingham, University Park, Nottingham NG7 2RD, UK\\
$^7$European Southern Observatory, Karl-Schwarzschild-Str.~2, 85748 Garching, Germany\\
$^8$Astronomy Centre, University of Sussex, Falmer, Brighton BN1 9QH, UK\\
$^9$SUPA, Institute for Astronomy, University of Edinburgh, Royal Observatory, Blackford Hill, Edinburgh, EH9 3HJ, UK. \\
$^{10}$School of Physics, Monash University, P.O. Box 27, VIC 3800, Australia\\
$^{11}$Department of Physics, Swiss Federal Institute of Technology (ETH-Z{\" u}rich), 8093 Z{\" u}rich, Switzerland\\
$^{12}$Institute for Computational Cosmology, Department of Physics, Durham University, South Road, Durham DH1 3LE, UK\\
$^{13}$Centre for Astrophysics and Supercomputing, Swinburne University, P.O. Box 218, Hawthorn, VIC 3122, Australia\\
$^{14}$Leiden University, P.O.~Box 9500, 2300 RA Leiden, The Netherlands\\
$^{15}$Institute of Cosmology and Gravitation (ICG), University of Portsmouth, Burnaby Road, Portsmouth PO1 3FX, UK\\
$^{16}$Jeremiah Horrocks Institute, University of Central Lancashire, Preston PR1 2HE, UK\\
$^{17}$Astronomy Unit, Queen Mary University London, Mile End Rd, London, E1 4NS, UK\\
$^{18}$Max Planck Institute for Nuclear Physics (MPIK), Saupfercheckweg 1, 69117 Heidelberg, Germany\\
}
\begin{document}

\date{}

\pagerange{\pageref{firstpage}--\pageref{lastpage}} \pubyear{2010}

\maketitle

\label{firstpage}

\begin{abstract}
We present an analysis of the properties of the lowest H$\alpha$-luminosity galaxies ($L_{H\alpha}\leq4\times10^{32}$ W; SFR$<0.02$ M$_\odot$yr$^{-1}$) in the Galaxy And Mass Assembly (GAMA) survey.  These galaxies make up the the rise above a Schechter function in the number density of systems seen at the faint end of the H$\alpha$ luminosity function.   Above our flux limit we find that these galaxies are principally composed of intrinsically low stellar mass systems (median stellar mass $=2.5\times10^8 M_\odot$) with only 5/90 having stellar masses $M>10^{10} M_\odot$.  The low SFR systems are found to exist predominantly in the lowest density environments (median density $\sim0.02$ galaxy Mpc$^{-2}$ with none in environments more dense than $\sim1.5$ galaxy Mpc$^{-2}$).  Their current specific star formation rates (SSFR; $-8.5 < $log(SSFR[yr$^{-1}] )<-12.$) are consistent with their having had a variety of star formation histories.  The low density environments of these galaxies demonstrates that such low-mass, star-forming systems can only remain as low-mass and forming stars if they reside sufficiently far from other galaxies to avoid being accreted, dispersed through tidal effects or having their gas reservoirs rendered ineffective through external processes.
\end{abstract}


\section{Introduction}
\label{sect:intro}

%Understanding the star formation history of the Universe is vital to any investigation of galaxy behaviour.  
Recent investigations into the evolutionary properties of galaxies have established that the distribution of galaxy colours exhibits a strong bimodality, separating red, quiescent early-type galaxies, with little ongoing star formation, from a blue sequence of star-forming spiral galaxies (e.g. \citealp{baldry06, blanton06, kauffmann03, driver06}).  This bimodality can be interpreted in terms of star formation history and stellar mass content (e.g. \citealt{tinsley68}) with the stars in the most massive galaxies having formed quickly a long time ago and less massive galaxies still forming stars today.
%, such that the characteristic mass of galaxies dominated by star formation is lower with decreasing redshift - a phenomenon known as `downsizing'  \citep{cowie96}.  However, it is unclear how the lowest mass galaxies fit into this picture.  Only galaxies with stellar masses above $10^9M_\odot$ have been able to be detected in the samples used to date to establish the relationship between star formation rate and mass with redshift (e.g. \citealt{juneau05,zheng07,mobasher09}).

The current star formation of a galaxy can be traced through its H$\alpha$ luminosity as this emission 
%arises from O-star formation, 
gives a direct probe of the young massive stellar population \citep{kennicutt98}.  The Galaxy And Mass Assembly survey (GAMA\footnote{http://www.gama-survey.org/}; \citealt{driver09,driver10}), has to date obtained optical spectra for $\sim120,000$ galaxies in the nearby Universe ($z<0.5$).  GAMA's deep spectroscopic observations ($r<19.8$; two magnitudes fainter than the Sloan Digital Sky Survey; SDSS; \citealt{york00}) and wide-area sky coverage (three 48 square-degree regions) provide an excellent sample with which to analyse the current star formation rates %(as described by the distribution of H$\alpha$ luminosities) 
of a large sample of galaxies, down to very low stellar masses.  Gunawardhana et al. (in prep.) have used early data from GAMA to establish that there is a significant increase in the number densities of H$\alpha$-emitting galaxies at very low luminosities (H$\alpha$ luminosities $<2.5\times10^{32}$ W, $H_0=100$ km s$^{-1}$ Mpc$^{-1}$) such that a Schechter Function is a poor fit at these luminosities.  This has also been observed by \cite{westra10} with the Smithsonian Hectospec Lensing Survey.  %The increase in number density suggests that galaxies with these low H$\alpha$ luminosities may not follow the bimodal star formation histories dependent on stellar mass observed in massive galaxies.

Characterising the faint-end of the H$\alpha$ luminosity function is important as both \cite{lee07} and \cite{bothwell09} have found that there is a broadening of the star formation distribution to extremely low star formation rates (SFR) in low-mass as well as high-mass galaxies from their analysis of H$\alpha$-derived SFRs for galaxies within 11 Mpc. The broadening of the star formation distribution in high mass galaxies is due to the cessation of star formation.  However, the broadening for low-mass galaxies is an open question with two potential resolutions: First, that the cause of this range is due to a change in the dominant physical process that regulates star formation. These galaxies are generally low mass and such galaxies have shallow potential wells.  They are therefore subject to many processes that are negligible in higher-mass systems. Second, the possibility that, at low luminosities, the star formation rate is not being traced as closely by the H$\alpha$ luminosity due to the existence of fewer young stars, meaning that the initial mass function (IMF) is not fully sampled in these systems. 
%\cite{lee09b} go on to examine how H$\alpha$ luminosity traces SFR in more detail. 
\cite{meurer09} and \cite{lee09b} find that the H$\alpha$ luminosity underestimates the SFR relative to the FUV luminosity in dwarf galaxies and that this could be the result of a steeper initial mass function (IMF) for galaxies with lower SFRs, for which there is growing theoretical  (e.g. \citealt{weidner05}) and observational evidence \citep{hoversten08,gunawardhana10}.

In this paper we investigate the characteristics of the low H$\alpha$-luminosity galaxies that comprise the upturn in the H$\alpha$ luminosity function. We investigate their stellar masses to determine whether the low H$\alpha$-luminosities are from low-mass galaxies with high specific star formation rates (SSFRs) or massive galaxies with low SSFRs and find that these are generally low-mass systems with high SSFRs.  We examine how much of the increase in number densities seen at low stellar masses ($<3\times10^{8}M_{\odot}$;  \citealt{baldry08}) in the galaxy stellar mass function is due to these systems and find that they do not contribute significantly.
%GAMA's deep spectroscopic data enables us to examine low H$\alpha$-luminosity systems to higher redshifts than previous large surveys.  We c.  
We also use the wide area of the GAMA survey to investigate 
%whether these galaxies follow the tight relationship of star formation rate with stellar mass (e.g. \citealt{brinchmann04, schiminovich07}) and 
what environment these low star forming galaxies are found in and conclude that they are only found in low density environments.
%, although at the lowest levels of specific star formation rate.

We describe the construction of the sample in Section \ref{sect:data}.  In Section \ref{sect:properties} we describe the observed properties, star formation rates and environmental density of the sample, and discuss our findings in Section \ref{sect:discussion}.  Throughout this paper we assume a Hubble constant of $H_0=70$ km s$^{-1}$ Mpc$^{-1}$ and an $\Omega_M=0.3$, $\Omega_\Lambda=0.7$ cosmology.  All magnitudes are given in the AB system.   We use a Salpeter IMF in our derivations of stellar masses and SFR for convenience but recognize that quantitative values may need to be scaled to a more realistic IMF.

\section{Data}
\label{sect:data}

%\subsection{GAMA}

The Galaxy And Mass Assembly (GAMA) survey is using wide-field survey facilities to study galaxy formation and evolution over a range of wavelengths. GAMA brings together data from the 3.9m Anglo-Australian Telescope (AAT), the United Kingdom InfraRed Telescope (UKIRT), the Very Large Telescope Survey Telescope (VST), the Visible and Infrared Survey Telescope for Astronomy (VISTA), the Australian Square Kilometre Array Pathfinder (ASKAP), and the Galaxy Evolution Explorer (GALEX) and Herschel space telescopes.  To date $\sim120,000$ galaxies in three 48 square degree regions of sky have been observed \citep{driver09,driver10}.   Here we use data from the first two years of GAMA optical spectroscopic observations undertaken with the dual-arm AAOmega spectrograph \citep{sharp06} at the AAT.  The GAMA input catalogue is described in detail in \citet{baldry09} and the spectroscopic tiling of those sources in \citet{robotham09}.  In summary, spectroscopic targets were initially selected from the SDSS Data Release 6 \citep{adelman-mccarthy08} to Galactic extinction-corrected, Petrosian magnitude limits of $r<19.4$ mag in 2 fields and $r<19.8$ mag in 1 field.  The input catalogue includes an explicit low surface brightness limit, to aid artefact removal, $\mu_{r,50}>15$ mag arcsec$^{-2}$, where $\mu_{r,50}$ is the effective $r$-band surface brightness within the half-light radius.  The targets were observed with the 580V and 385R AAOmega gratings giving an observed wavelength range of $\sim3700 - 8900\rm{\AA}$ and spectral resolution of $3.2\rm{\AA}$ FWHM.  H$\alpha$ is within the observed wavelength range to redshifts $z\sim0.35$.  The spectra are extracted with \texttt{2DFDR} which undertakes a first-order curvature correction and flux calibration in order to accurately splice the spectra from the blue- and red-arms of the spectrograph.  The spectra are then sky-subtracted following \cite{sharp10b} and redshifts determined with  \texttt{RUNZ} \citep{saunders04}.  In the first two years of GAMA data there are 
%98,872 
81,274 GAMA redshifts with quality values $Q\geq3$ (i.e. regarded as a secure redshift).  

%To measure the H$\alpha$ emission necessary for this analysis, 
The standard strong optical emission lines are measured from each flux-calibrated spectrum.  This is done assuming a single Gaussian approximation, fitting for a common velocity and line-width within an adjacent set of lines (e.g. H$\beta$ and the [OIII] doublet and separately, H$\alpha$ and the [NII] doublet) whilst simultaneously fitting the continuum local to the set of lines with a straight-line fit.  We measure the limit to which our H$\alpha$ fluxes are robust from a self-consistent error estimation of emission line fits \citep{acosta-pulido96}.  Since the ratios of the fluxes of emission-line doublets are purely set by quantum mechanics, we use measurements of the [OIII] and [NII] doublet fluxes to measure the flux at which the ratio is no longer accurate to within 10 percent.  For a single line this then translates to a flux limit of $3\times10^{-16}$erg s$^{-1}$cm$^{-2}$.  We determine the limit to which our H$\alpha$ fluxes are complete (i.e. $1\times10^{-15}$erg s$^{-1}$cm$^{-2}$) from the number counts of our measurements.   We make no attempt to construct a complete, volume or mass-limited sample as in order to obtain a reasonable range in luminosity or redshift ends up excluding the low H$\alpha$ flux systems that we are studying.  Instead we take advantage of the fact that we can detect systems to these faint levels to explore the nature of some of the most extreme systems in the GAMA sample.

%The H$\alpha$ flux limit is $3\times10^{-18}$ergs s$^{-1}$cm$^{-2}$ and the measurements are complete to $1\times10^{19}$ergs s$^{-1}$cm$^{-2}$.  
%In later GAMA data releases spectra fitted with the GANDALF stellar population fitting routine that fits galaxy templates to both the underlying continuum and the emission lines \citep{sarzi06} will also be available for this analysis.

%Corrections for the underlying stellar population, dust obscuration and fibre aperture effects are then applied to these measurements (see \ref{Madusa10} for further details...).

%The 1D spectra are fitted with Gaussian functions to obtain intensities and equivalent widths (EW) for a range of emission lines (OII - H$\alpha$).  With a detection limit of H$\alpha \sim$, it is possible to obtain fits for XX galaxies.

A significant increase in the number densities of H$\alpha$-emitting galaxies at very low luminosities is observed in the lowest redshift bin of the GAMA H$\alpha$ luminosity function ($z\leq0.13$; Gunawardhana, et al., in prep.).  We therefore focus on galaxies within this redshift range.  There are 13,228 GAMA galaxies with $Q\geq3$ and $0.002<z<0.13$ (henceforth referred to as the Whole sample for the purposes of this analysis) and of these, 9,623 galaxies have H$\alpha$ fluxes above our flux limit.  We also require that H$\beta$ is measured as it is necessary for dust corrections described later.  As H$\beta$ is a significantly weaker emission line, we only require that H$\beta$ is measured, rather than imposing a flux limit, in order not to discard weak line systems \citep{cid-fernandes10}.  As we are studying the star-forming properties of this sample, we remove 707 Active Galactic Nuclei (AGN) following the diagnostic relationships from \citet{kewley01}.  These relationships rely on four emission lines.  If we do not have measurements for all four then we use the two-line diagnostics.  If we do not have both lines for the two-line diagnostic then the galaxy is classified as `uncertain', but remains in the sample to prevent discarding weak-line systems.  Using the \citet{kewley01} diagnostics only removes 4 galaxies from the final low-H$\alpha$ luminosity sample and does not change any of our conclusions.
%We also note that most low luminosity galaxies are more likely to host central compact stellar nuclei than a central AGN \citep{cote06} and neither \cite{kauffmann03} nor \cite{haines07} find  optically-identified AGN with $M_r>-18$ in similarly selected SDSS samples.  
This leaves a sample of 8,916 galaxies.

%\subsection{Distances}

%We are primarily interested in the lowest luminosity galaxies in the lowest redshift bin ($z\leq0.13$) of the GAMA H$\alpha$ luminosity function (Gunawardhana, et al., in prep.). 
At redshifts below $z\sim0.02$
%(Figure \ref{z_lha_all}) 
peculiar velocities are a significant fraction of the measured recessional velocity.  This can lead to distance errors of a factor of 2 or more when assuming pure Hubble flow (e.g. \citealt{masters04}).  It is therefore important to include local variations from the cosmic microwave background (CMB) frame.  We use the GAMA peculiar velocity corrections which are calculated by applying the \citet{tonry00} multi-attractor flow model which provides a parametric model for the local velocity field, from $z\sim0.002$ (below which stellar contamination arises) to $z\sim0.02$. From $0.02<z_{CMB}<0.03$ the peculiar velocity corrections are linearly tapered to zero by maximising the probability density of the corrections, and above $z\sim0.03$ a pure $z_{CMB}$ is invoked.

\begin{figure}
\begin{center}	
\resizebox{20pc}{!}{
\includegraphics{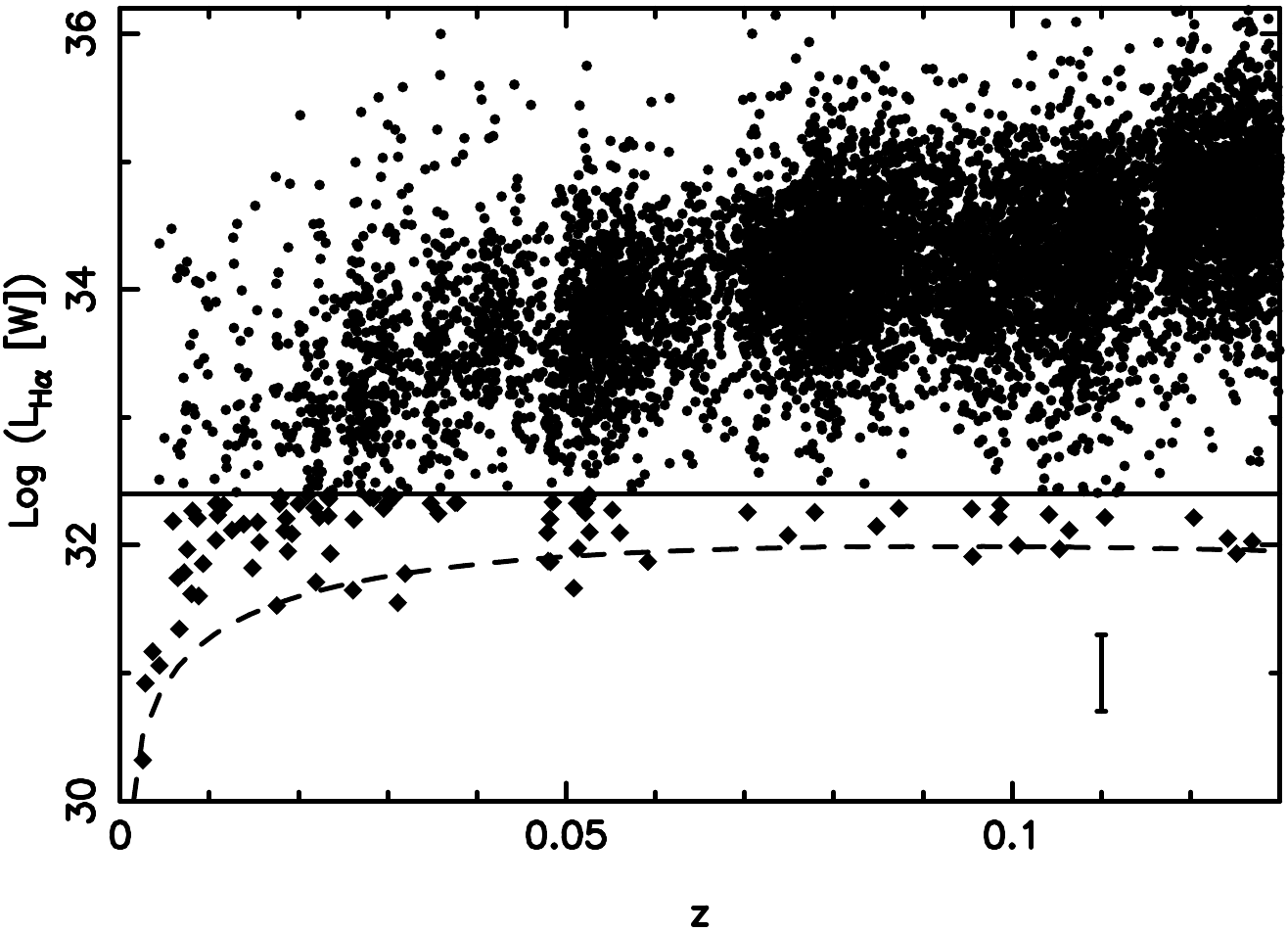}
}
\end{center}
\caption{Distribution of H$\alpha$ luminosity, L$_{H\alpha}$, with redshift for the Main sample (star-forming galaxies with $0.002<z<0.13$).  Gaps in the distribution are primarily due to large-scale structure at the low redshift end, and are due to the H$\alpha$ line passing through the Fraunhofer OH atmospheric absorption bands at higher redshifts.  The solid line indicates the cut made to select the lowest H$\alpha$ luminosity galaxies.  The flux limit of our observations is indicated by the dashed line.  The median $1\sigma$ error on the H$\alpha$ luminosity is given in the bottom right-hand corner.}
\label{z_lha_all}
\end{figure}

%\subsection{Absolute Magnitudes}

We calculate intrinsic galaxy luminosities using GAMA $g, r, i$ magnitudes measured in $r$-band-defined elliptical Kron apertures \citep{hill10}.  These magnitudes are corrected for Galactic extinction according to the dust maps of \cite{schlegel98} and $K$-corrected to a redshift $z=0$ using the  \textsc{KCORRECT V4\_1\_4} code of \cite{blanton07} such that $M_x=m_x - 5 log_{10}(D_L[Mpc]) - 25 - K_x$.   $r$-band magnitudes re-computed for GAMA are not available for 418 galaxies.  These missing magnitudes are generally either due to a SExtractor detection failure or a particularly low surface brightness source \citep{hill10}.  This leaves a sample of 8,666 galaxies (henceforth referred to as the Main sample).

\subsection{Low H$\alpha$ Luminosity Sample}

In order to calculate the H${\alpha}$ luminosity we first correct the observed equivalent width, EW(H${\alpha}$), for the effects of stellar absorption within each galaxy, aperture effects from the $2^{\prime\prime}$ fibres, and intrinsic dust obscuration.  We adopt the method of \citet{hopkins03} for calculating total H${\alpha}$ luminosities from SDSS fibre spectroscopy:
%Following the emission-line fitting, removing AGN and correcting the galaxies to a consistent distance scale; we select the galaxies with the lowest H${\alpha}$ luminosities, $L_{H{\alpha}}<$. 
\begin{eqnarray}
L_{H\alpha} (W) & = & (EW+EW_c)10^{-0.4(M_r-34.10)} \nonumber \\
& & \times \frac{3\times10^{18}}{[6564.61(1+z)]^2}
% \nonumber \\ &&
 \times \left(\frac{S_{H\alpha}/S_{H\beta}}{2.86}\right)^{2.36}
\end{eqnarray}
A constant correction, $EW_c$, is added to the Balmer emission lines to correct the measured EW for internal stellar absorption.  \citet{hopkins03} argued that a correction of $EW_c=1.3\rm{\AA}$ was sufficient. However, an analysis of GAMA data indicates that results are quantitatively and qualitatively similar if the correction is between $0.7$ and $1.3$ \citep{gunawardhana10} and we therefore use the smaller correction in all analyses, $EW_c=0.7\rm{\AA}$.  We correct for the difference in aperture size between the AAT fibre ($2^{\prime\prime}$ diameter) and the galaxy itself by using the absolute, extinction-corrected, $r$-band  magnitude, $M_r$ \citep{hill10}, to approximate the continuum at the wavelength of H$\alpha$, with the implicit assumption that the H$\alpha$ emission is traced by the continuum emission \citep{hopkins03,brinchmann04}.  The factor of $3\times10^{18}/{[6564.61(1+z)]^2}$ converts from units of W Hz$^{-1}$ to W \rm{\AA}$^{-1}$, for the wavelength of H$\alpha$ in a vacuum.  H$\alpha$ emission can be heavily attenuated by dust. We therefore correct for dust extinction intrinsic to the galaxy using the Balmer Decrement, assuming an intrinsic Case B recombination ratio of 2.86 for H$\alpha$/H$\beta$, (S$_{\rm{H}\alpha}$ and S$_{\rm{H}\beta}$ are the stellar absorption corrected line fluxes, see \citealt{hopkins03}) and the \cite{cardelli89} Milky Way dust correction for emission lines, giving an exponent of  2.36.  This gives a median correction factor of no obscuration correction for these galaxies, consistent with their low luminosities.
%, rather than the 2.114 used in \citet{hopkins03}.

From the Main sample of 8,693 galaxies
%We make these measurements for galaxies with H${\alpha}$ and H${\beta}$ fluxes which are at least twice as large as the errors on the fluxes. %leaving a sample of 22,511 galaxies.  
%From these robust measurements 
we select the least H${\alpha}$ luminous galaxies above the flux limit of $3\times10^{-16}$ergs s$^{-1}$cm$^{-2}$ up to the luminosity at which the upturn in the H$\alpha$ luminosity function is observed (\citealt{westra10}, Gunawardhana et al., in prep.): $2\times10^{30}\leq L_{H\alpha}\leq4\times10^{32}$ W.  The upper limit is equivalent to a star formation rate of $0.02$ M$_\odot$yr$^{-1}$, significantly less than the rate observed in the Milky Way of $\sim1$ M$_\odot$yr$^{-1}$ \citep{robitaille10}.
%(XX LAST BIN or bottom 100 or LHalpha or XX)

As these galaxies are approaching the edge of the GAMA selection criteria, SDSS thumbnail images and AAT spectra were compiled for each galaxy and visually inspected to check for discrepancies.  The GAMA catalogue is derived from an automated procedure applied to SDSS and UKIDSS \citep{baldry09}. In certain cases sources were visually checked and given the classification \textsc{vis\_class} $=3$ if the object, such as an HII region, was part of a larger system. These are not included in our sample. In addition, we rejected a further 2 sources with stars less than $2^{\prime\prime}$ away (as the spectra are contaminated by those stars) and 3 sources which were HII regions not previously identified.
%In the year 2 data, at redshifts $z>0.002$ there remains some stellar contamination and HII regions misidentified as galaxies in SDSS due to over-de-blending \citep{baldry09,hill10}.  We remove 2 galaxies with stars $<2^{\prime\prime}$ away; and 3 galaxies which are clearly HII regions of a larger system.  The HII regions are all now classified as VIS$\_$CLASS=3 i.e. not a target due to not being the main part of a galaxy, \citep{baldry09} in the latest GAMA parent catalogue (I. Baldry, private comm.).  
This leaves a sample of 90 galaxies with $0.002<z<0.13$ and SFR $<0.02$ M$_\odot$yr$^{-1}$ (henceforth referred to as the low-$L_{H\alpha}$ sample).   The redshift distribution of the low-$L_{H\alpha}$ sample with respect to the Main sample is illustrated in Figure \ref{z_lha_all}.  The effects of our H$\alpha$ flux limit and H$\beta$ selection are also illustrated.   

We describe the physical and star formation properties of these galaxies in more detail in the next section.

\section{Physical and Star Formation Properties}
\label{sect:properties}

\begin{figure*}
\begin{center}	
\resizebox{20pc}{!}{ \includegraphics{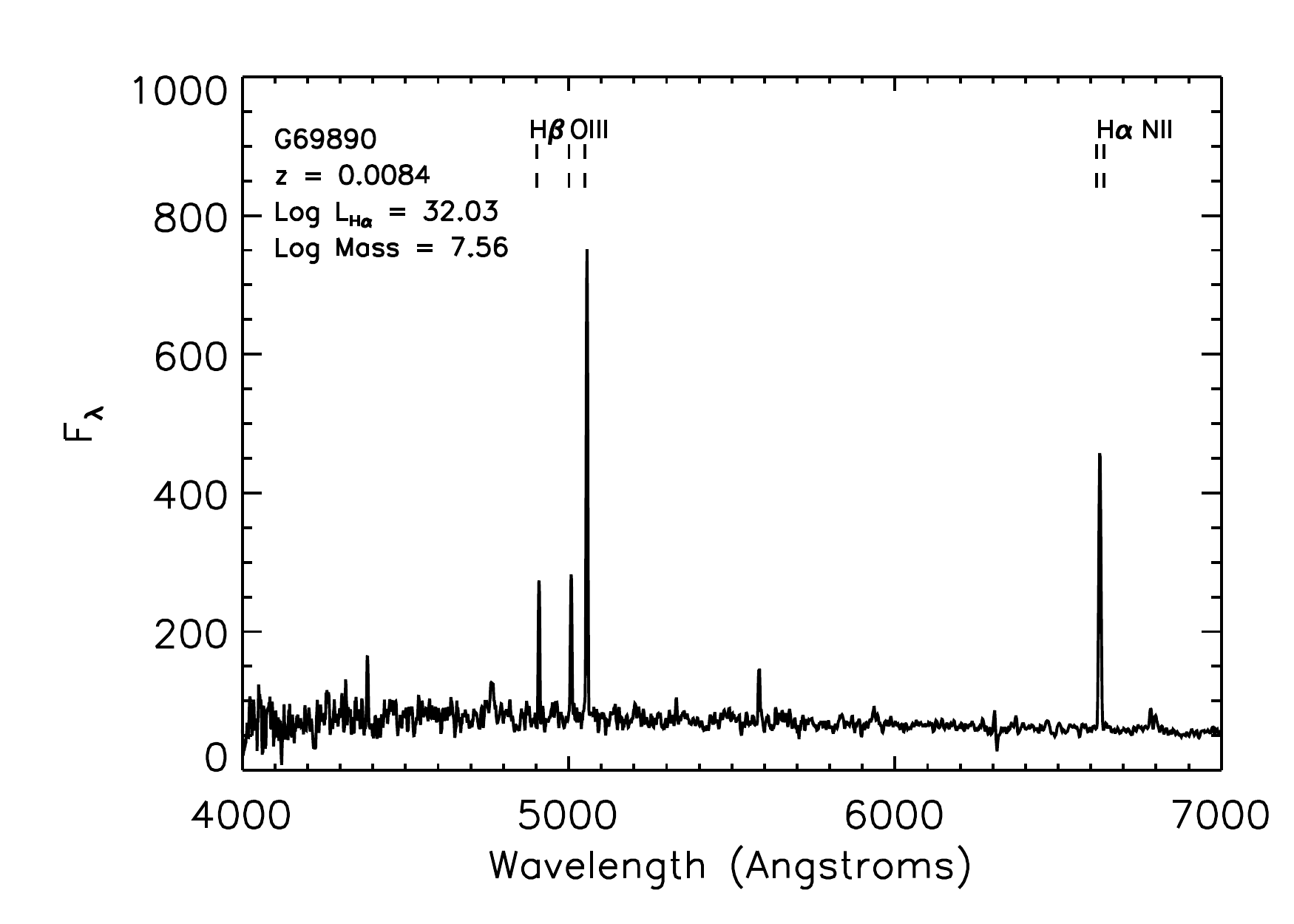}}
\resizebox{12pc}{!}{ \includegraphics{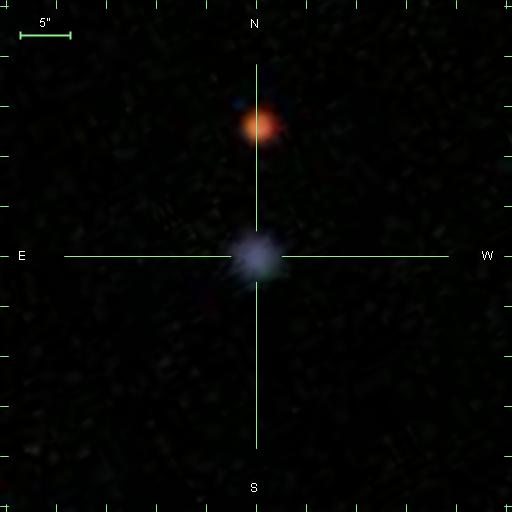} }
\resizebox{20pc}{!}{  \includegraphics{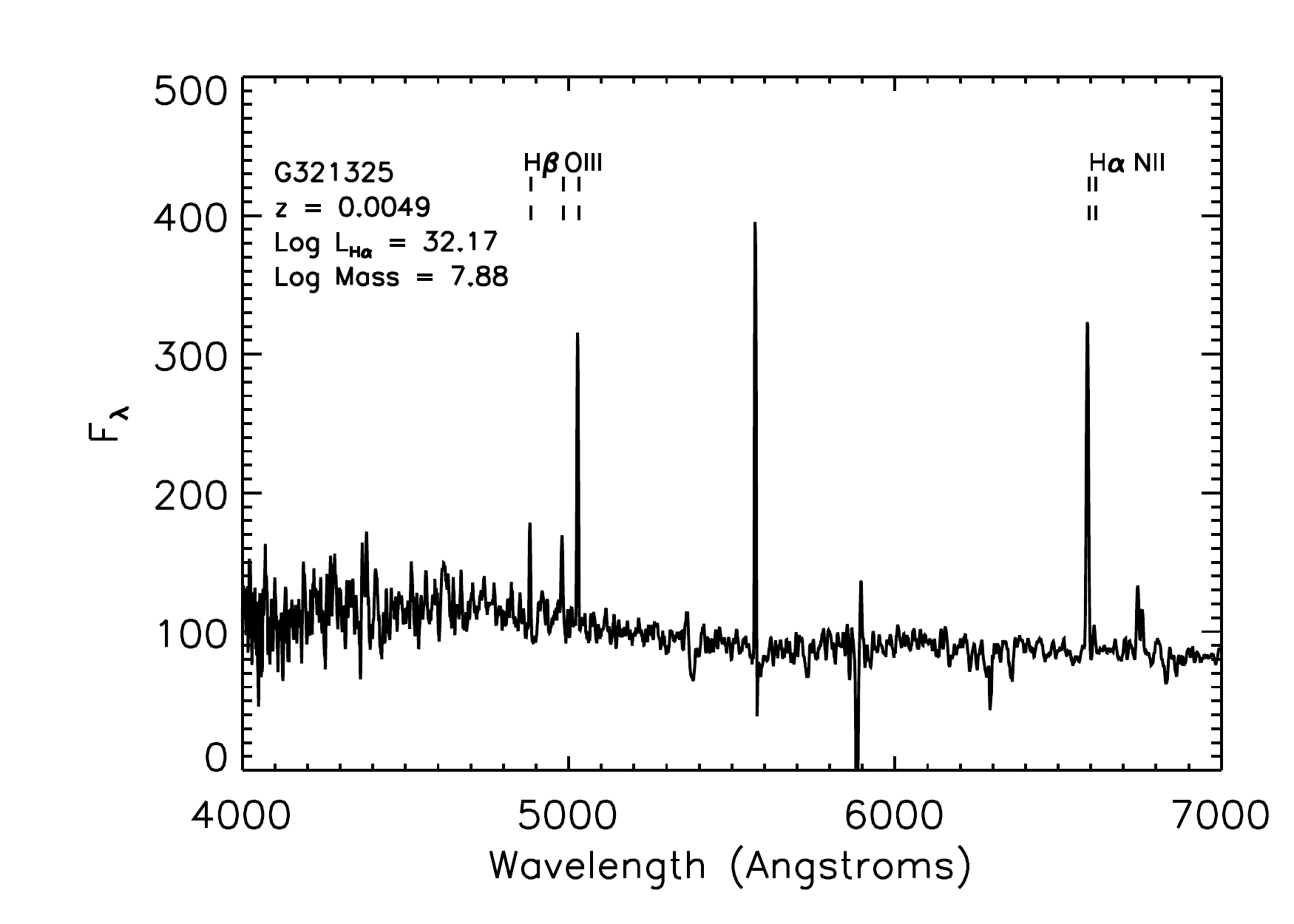} } 
\resizebox{12pc}{!}{ \includegraphics{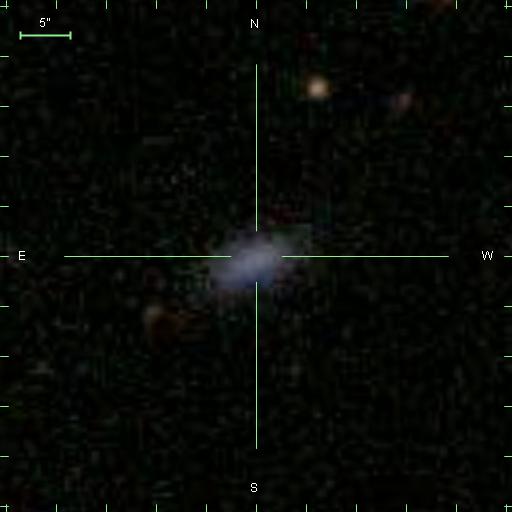} }
%\resizebox{22pc}{!}{  \includegraphics{../G92904} } 
%\resizebox{13pc}{!}{ \includegraphics{G00092904} }

\end{center}
\caption{AAOmega spectra (with prominent emission lines marked) and SDSS multi-colour images of (in each panel) GAMA galaxies: G69890 and G321325.  Names, redshifts, H$\alpha$ luminosities (W) and stellar masses (M$_\odot$) are given in each panel.  The scale bar given in the images corresponds to $5^{\prime\prime}$, at the redshifts of these systems that is equivalent to 0.8 kpc (G69890) and 0.5 kpc (G321325)}
\label{example}
\end{figure*}

In order to understand the physical and star formation characteristics of these galaxies in more detail, we examine their optical spectra and multi-colour images. We show the optical spectra (from the AAT) and multi-colour images (from the SDSS) for two galaxies, illustrative of the low-$L_{H\alpha}$ sample in Figure \ref{example}.  These clearly indicate that our selection criteria of low-H${\alpha}$ emission identifies diffuse, blue, emission-line galaxies.  
%More quantitative properties are described below.

\begin{figure}
\begin{center}	
\resizebox{20pc}{!}{
\includegraphics{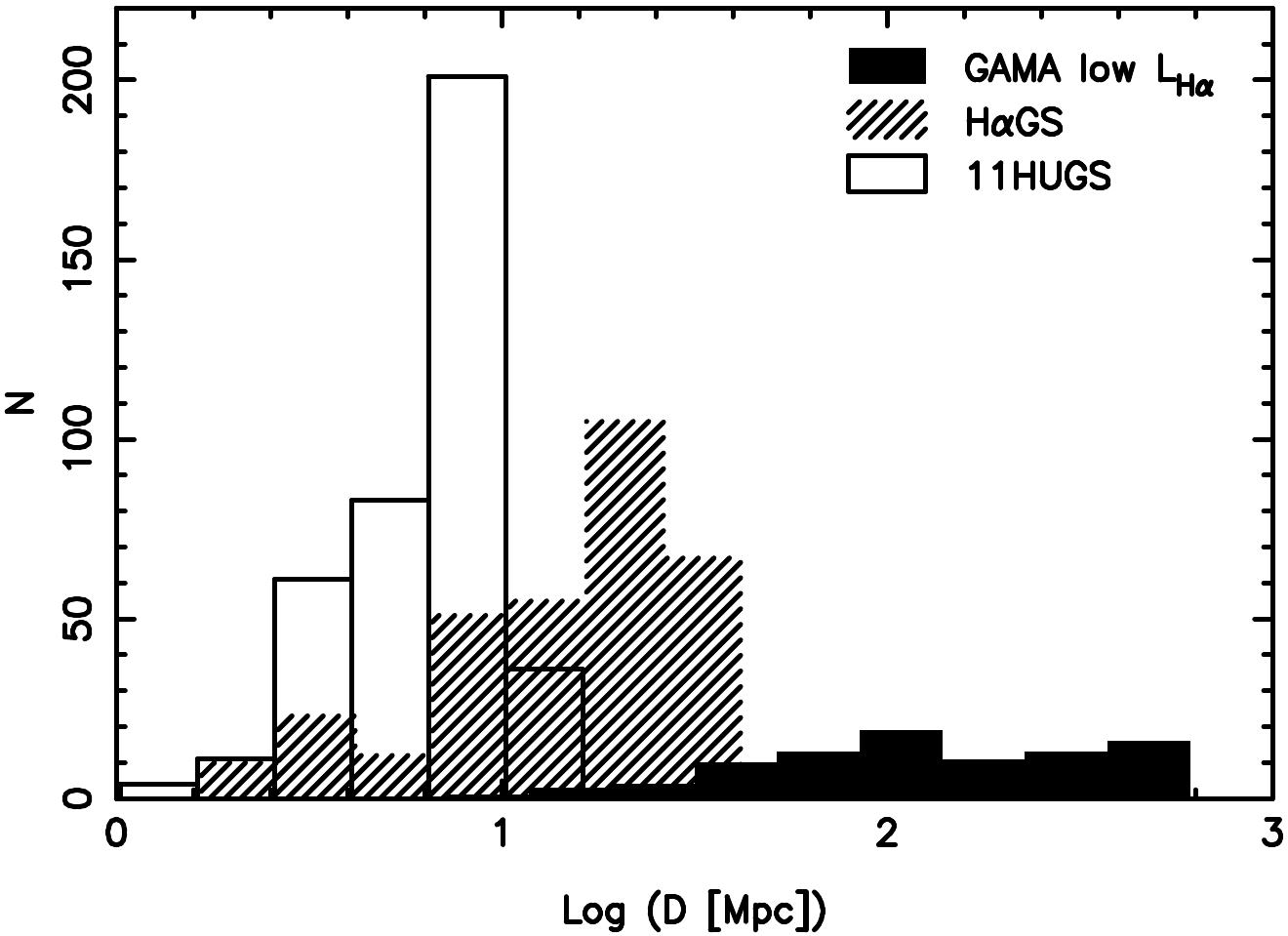}
}
\resizebox{20pc}{!}{
\includegraphics{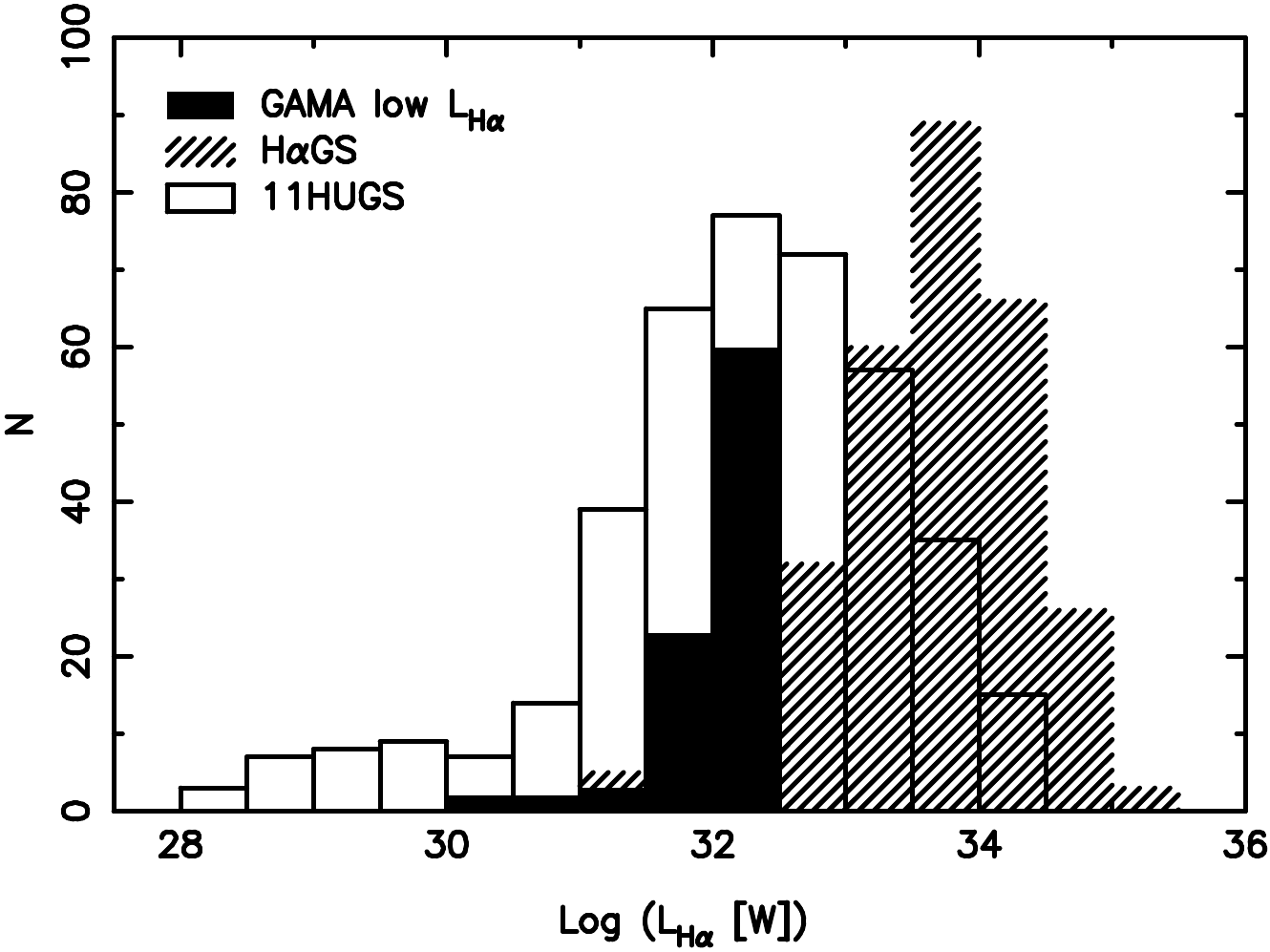}
}

%\resizebox{18pc}{!}{
%\includegraphics{../bmag_histo}
%}

%\resizebox{18pc}{!}{
%\includegraphics{../rad_mag}
%}

\end{center}
\caption{Distribution of low-L$_{\rm{H}\alpha}$ sample properties compared to other H$\alpha$ imaging surveys of the Local Volume, as a function of (a) Distance, D and (b) H$\alpha$ luminosity, $L_{H\alpha}$.}
%; and (c) absolute $B$-band magnitudes, $M_B$.} 
%GAMA data is indicated by the filled histograms, 11HUGS by the open histograms and H$\alpha$GS by the hatched histograms.}
\label{histos}
\end{figure}

In Figure \ref{histos} we compare the distribution of distances and H$\alpha$ luminosities for the low-$L_{H\alpha}$ sample with galaxies from H$\alpha$ imaging surveys of the Local Volume: the 11 Mpc H$\alpha$ and Ultraviolet Galaxy Survey   (11HUGS; \citealt{kennicutt08}) and the H$\alpha$ Galaxy survey (H$\alpha$GS; \citealt{james04}) which extends to $z\sim0.01$. Figure \ref{histos} (a) emphasises the much larger distance scale probed by the GAMA sample.  Our sample is the most distant of the three and Figure \ref{histos} (b) shows that we find similarly low H$\alpha$ luminosities to the H$\alpha$GS sample and to the majority of the significantly nearer 11HUGS sample.  %H$\alpha$GS is selected from the Uppsala Galaxy Catalogue with Hubble types S0/a and later, although both H$\alpha$GS and 11HUGS are dominated by dwarf irregular galaxies. 
This suggests that these are some of the most distant, low $H\alpha$ luminosity galaxies studied. 

\begin{figure}
\begin{center}	
\resizebox{20pc}{!}{
\includegraphics{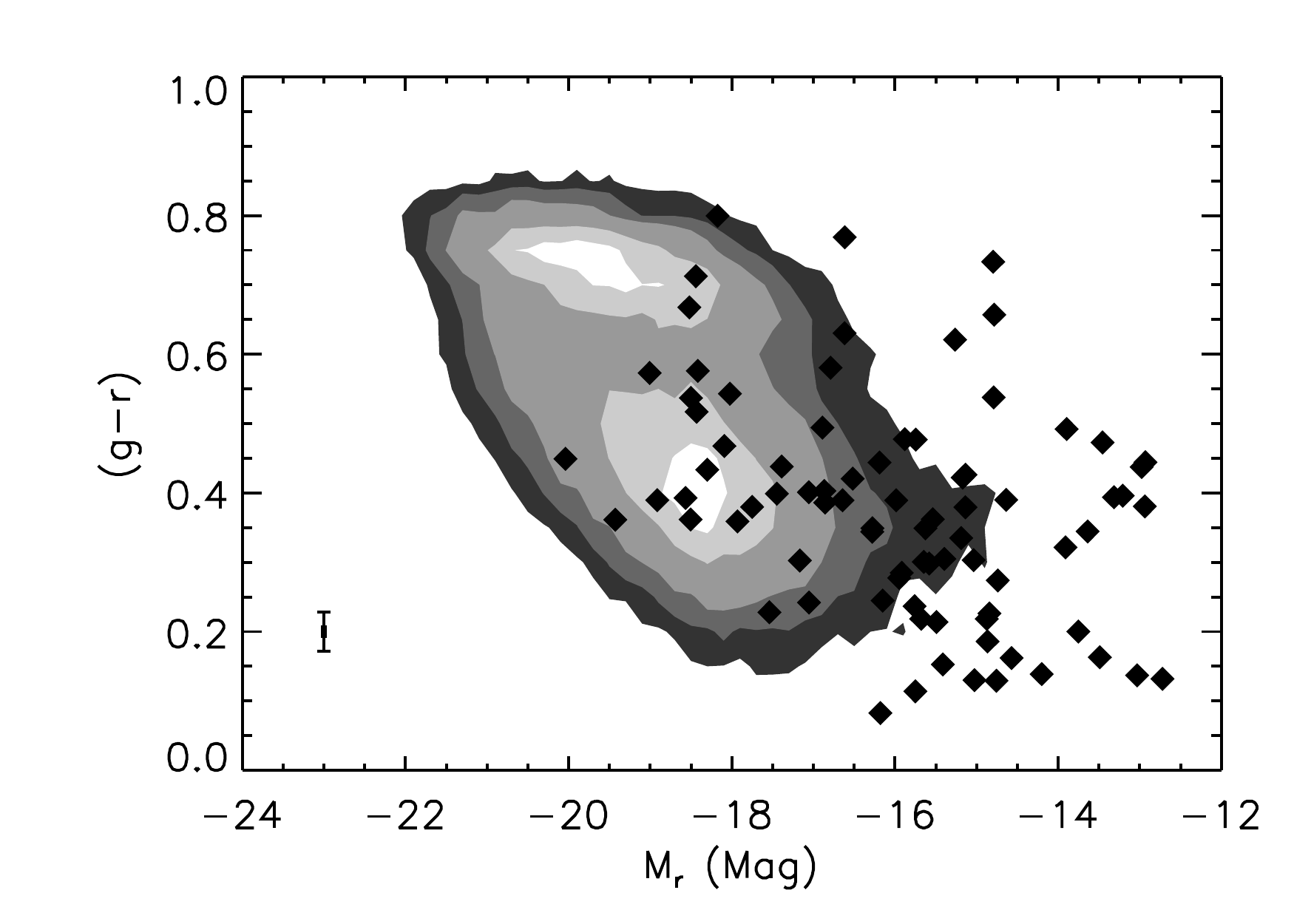}
}
\end{center}
\caption{$(g-r)$ colour-magnitude relation.  The contours indicate the number density of low-redshift NYU-VAGC galaxies.  The low L$_{\rm{H}\alpha}$ sample are indicated by the diamonds and can be seen to generally lie on the blue cloud (i.e. $g-r<0.55$), with lower luminosities and blue colours.  Those in the red sequence are the most massive of the GAMA sample.  The median $1\sigma$ errors on the magnitudes and colours are given in the bottom left-hand corner.}

%Distribution of $^{0.1}(g-r)$ colour with absolute $r$-band magnitudes, $^{0.1}M_r$.  GAMA data are indicated by the red points, NYU-VAGC data by the black points.}
\label{gr_colour_mag}
\end{figure}

We compare the colours of these galaxies to those from the New York University (NYU) re-analysis of SDSS low redshift data: the NYU Value-Added Galaxy Catalogue  (NYU-VAGC DR4; $10<$Distance (Mpc$/h) <150$; \citealt{blanton06}) in Figure \ref{gr_colour_mag}.  The GAMA data are clearly sampling lower luminosities than the majority of the NYU-VAGC low-redshift galaxies.  Most are of a similar colour to the `blue cloud' of star-forming galaxies (i.e. $g-r<0.55$) showing that these low-H$\alpha$-luminosity galaxies follow similar colour relationships to more massive star forming galaxies.  There are a few galaxies on the red sequence and these are some of the more massive galaxies of our sample.

%pThis is due to the flux limit of our parent sample and the mass-SFR relationship which means that higher mass galaxies have lower star formation rates and are therefore redder. 
%Visual inspection does not reveal significantly different characteristics for these galaxies.  

% shows the colour-magnitude relation for these galaxies, indicating that the mean  colour of these galaxies is $^{0.1}(g-i)=0.57$.}

\begin{figure}
\begin{center}	
\resizebox{20pc}{!}{
\includegraphics{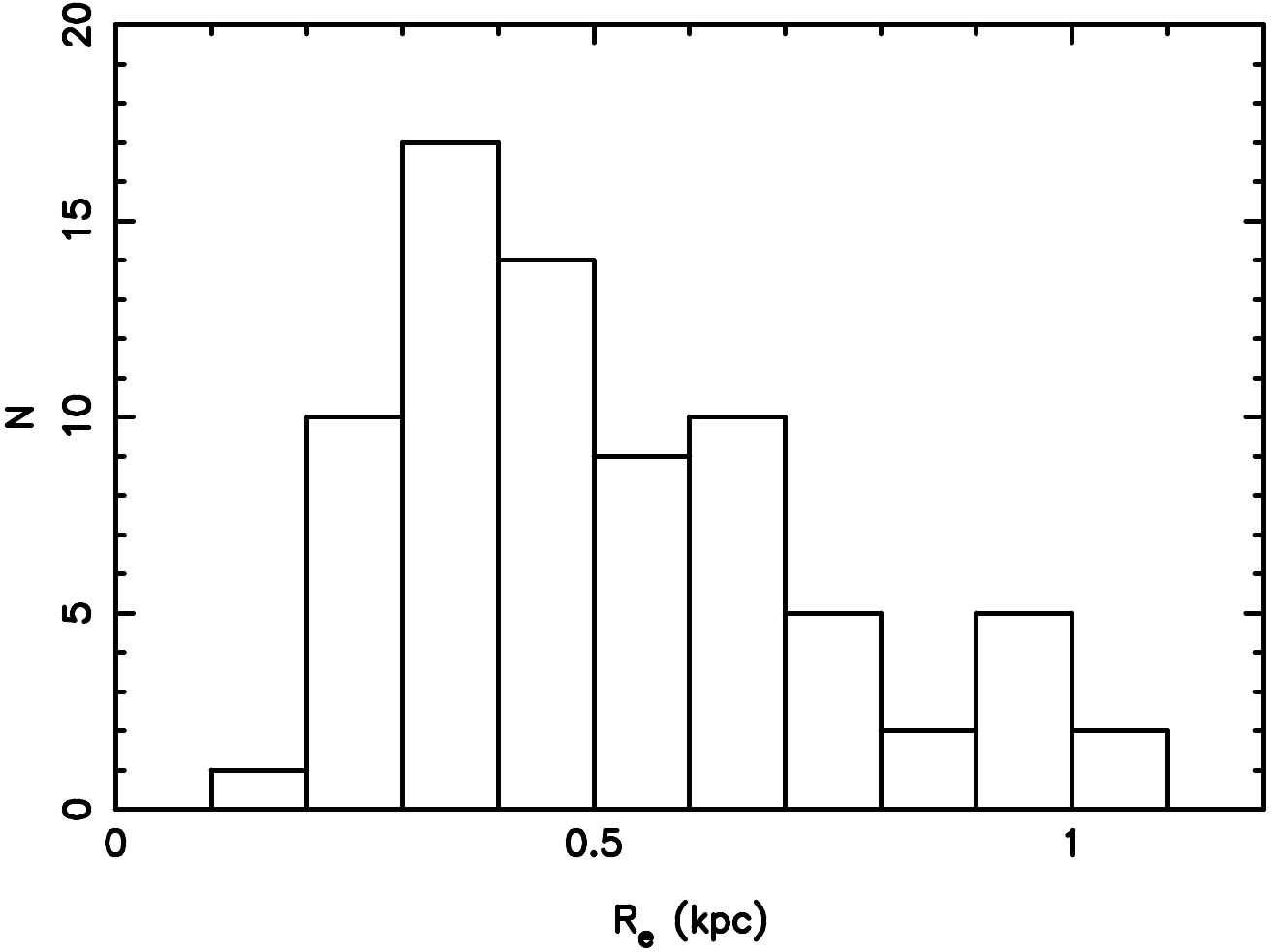}
}
\end{center}
\caption{ Histogram of Sersic $r$-band effective radii, $R_e$.  The low L$_{\rm{H}\alpha}$ sample all have low half-light radii $\leq1.2$ kpc.}

%Distribution of $^{0.1}(g-r)$ colour with absolute $r$-band magnitudes, $^{0.1}M_r$.  GAMA data are indicated by the red points, NYU-VAGC data by the black points.}
\label{rad_histo}
\end{figure}

We examine the size of these galaxies in Figure~\ref{rad_histo} using the effective radius (containing half of the flux) from a Sersic fit to the $r$-band data.  These galaxies are remarkably compact with all having half-light radii $\leq1.2$ kpc.  This result is unchanged if we use the SDSS  r$_{50}$ radius which measures the radius containing 50 per cent of the Petrosian flux.

%\subsection{Stellar Mass}

\begin{figure}
\begin{center}	
\resizebox{20pc}{!}{
\includegraphics{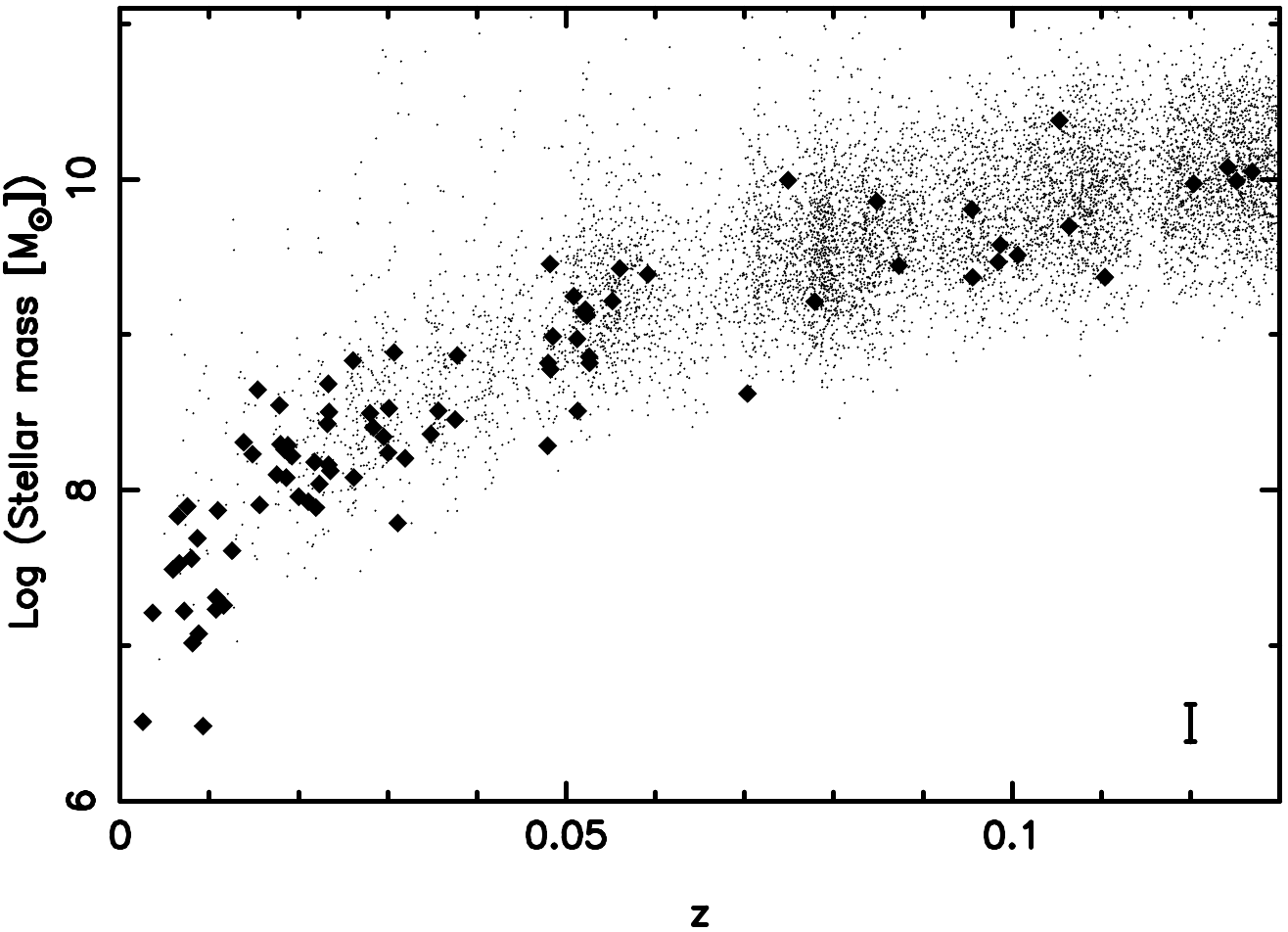}
}
\end{center}
\caption{Distribution of stellar mass with redshift for the Main sample (star-forming galaxies with $0.002<z<0.13$).  The diamonds indicate the low L$_{\rm{H}\alpha}$ sample.  The formal uncertainty on the stellar masses is $0.12$ dex and is indicated by the error bar in the bottom-right corner.}
\label{z_mass_all}
\end{figure}

Of primary interest to this analysis are the masses of these galaxies.  We use total stellar masses calculated from fits to the $u,g,r,i,z$ spectral energy distribution using the \cite{bruzual03} stellar population models, assuming a \cite{chabrier01} IMF (Taylor et al. in prep). %and a \cite{calzetti00} dust obscuration law.  
We adjust these to the Salpeter IMF used here by adding $\sim0.2$ dex.  The distribution of stellar masses with redshift with respect to the Main sample is illustrated in Figure~\ref{z_mass_all}.  It is clear that our low L$_{\rm{H}\alpha}$ sample are very low mass at low redshifts but have increasingly higher masses at higher redshifts.  In general we find that these galaxies have low stellar masses, however five have stellar masses $M\geq10^{10}M_\odot$.  The median stellar mass of the low-H$\alpha$ luminosity sample is $2.5\times10^8 M_\odot$, roughly that of the Small Magellanic Cloud.  This is a result of our flux limit, as will be seen in the next section.
% and the number density of high mass galaxies, which is significantly lower than that seen at the faint-end of the H$\alpha$ luminosity function.

\begin{figure}
\begin{center}	
\resizebox{20pc}{!}{
\includegraphics{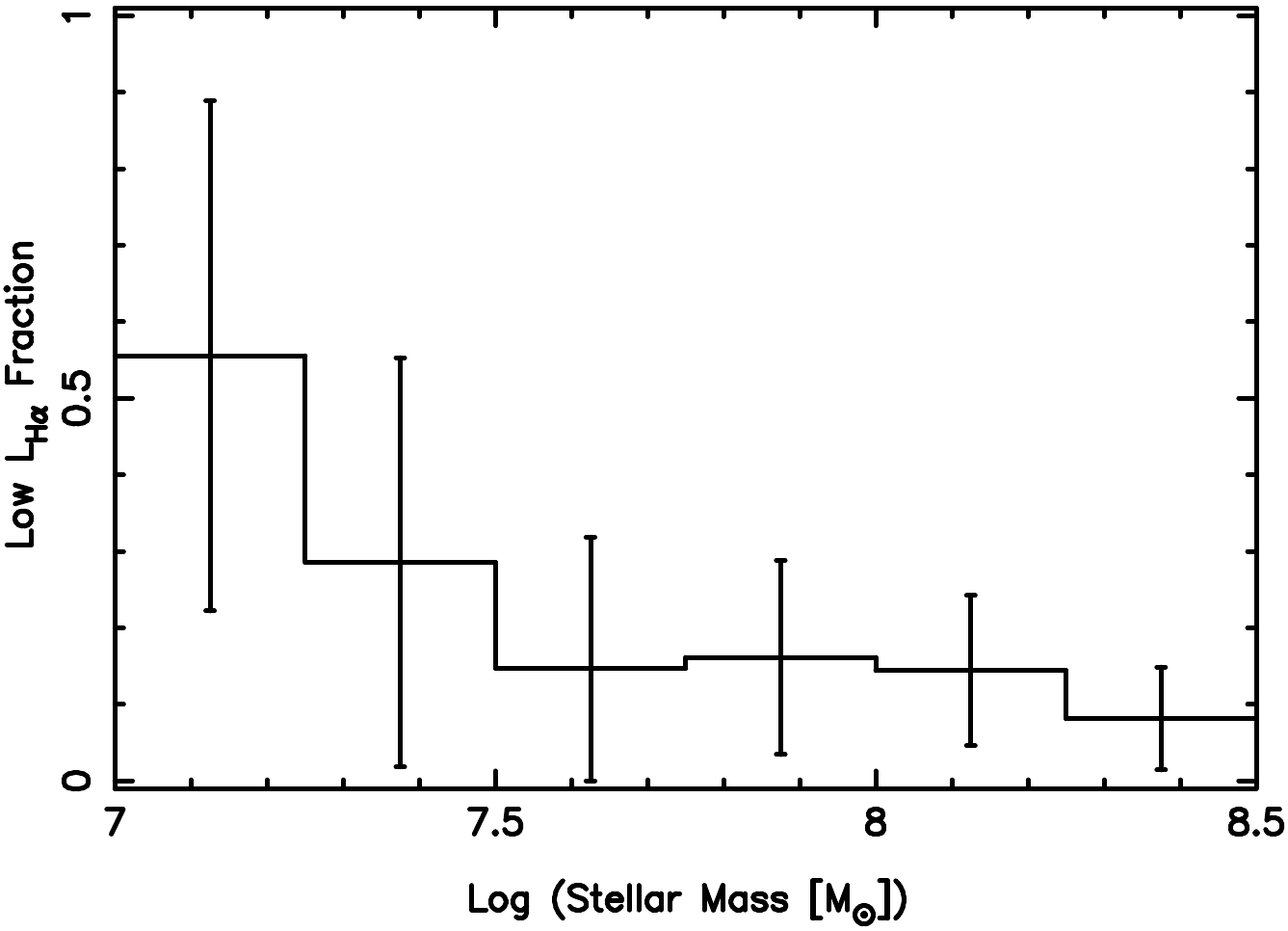}
}
\end{center}
\caption{Fraction of the low $L_{H\alpha}$ sample as a function of the Whole sample (all galaxies with $0.002<z<0.13$), in bins of stellar mass.  The error bars indicate poisson errors on the number of galaxies in each stellar mass bin.}
\label{mass_histo}
\end{figure}

%(relevant for $g-i$ colours between $0.4 < ^{0.1}(g - i) < 1.8$, which is applicable to this sample.  
%The distribution in mass of the low-$L_{H\alpha}$ sample with respect to the main sample is illustrated in Figure \ref{ssfr_mass}.  
The galaxies that make up the upturn in the H$\alpha$ luminosity function are of similar stellar mass to those that form the increase in number density seen in the galaxy stellar mass function at stellar masses $\leq3\times10^{8}M_{\odot}$  \citep{baldry08}. To examine the contribution of these low star formation galaxies to the upturn in the stellar-mass function we show the fraction of galaxies classified as having low-H$\alpha$ luminosity as a function of stellar mass $<3\times10^{8}M_{\odot}$ in Figure \ref{mass_histo}.   As stellar mass decreases the low-H$\alpha$ luminosity sample are an increasing fraction of the Whole galaxy population and the low star formation galaxies form the largest fraction of the mass function at stellar masses $\sim10^{7}M_{\odot}$. However, the small numbers of low-H$\alpha$ luminosity galaxies mean that this is not statistically significant.  At higher masses galaxies with higher star formation rates dominate the mass function.  %At lower stellar masses the number of star-forming galaxies appear to be falling off,  however visual inspection of those galaxies determined to be passive with M $\leq10^{9}M_{\odot}$ reveals a population of blue galaxies whose emission lines are weaker than our detection limit.  This is consistent with  H$\alpha$ imaging surveys of the local Universe that find all galaxies show signs of H$\alpha$ emission down to their detection limits (e.g. \citealt{james04,kennicutt08}).  
The upturn observed in the H$\alpha$ luminosity function is not cleanly responsible for the upturn observed in the galaxy stellar mass density at the lowest stellar masses.
%The low stellar masses mean that these galaxies certainly contribute towards the .   
%The upturn in number densities at low stellar masses is a result of decreased feedback efficiency with respect to higher stellar mass galaxies.  
We investigate the star formation histories of these galaxies with respect to their stellar masses in the next section.

\subsection{Star Formation Histories}

%\begin{figure}
%\begin{center}	
%\resizebox{20pc}{!}{
%\includegraphics{../mass_td}
%}
%\end{center}
%\caption{Star Formation Rate (SFR) as a function of stellar mass for the main sample. The low H$\alpha$-luminosity sample is indicated by the diamonds and can be seen to lie at the low-SFR, low-mass-edge of the relationship.}
%\label{sfr_mass}
%\end{figure}

%In Figure \ref{sfr_mass} we compare the SFR of these galaxies to their current stellar masses.  The SFR-mass relationship can clearly be seen. The lowest H$\alpha$-luminosity sample form the low-mass, low-SFR tail to this distribution.

%XX It is clear that the values we measure are consistent with these galaxies having formed their stellar mass through constant star formation over a Hubble time. XX

%\subsection{Specific Star Formation Rate (SSFR)}

In \S \ref{sect:intro} we stressed the importance of star formation history (SFH) on the observed properties of galaxies. To first order, we can characterise the star formation history using the specific star formation rate (SSFR).  It is defined as the ratio of the current Star Formation Rate (SFR) to the stellar mass formed to date and is closely related to Scalo's stellar birthrate parameter, $b$ \citep{kennicutt94}.  The SSFR gives a measure of whether the current star formation activity is typical of the past star formation activity (i.e. a constant star formation rate; Log(SSFR) $\sim-10$, $b\sim1$), the current star formation rate is elevated over the past (i.e. star-bursting; Log(SSFR) $>-10$, $b>>1$) or is reduced with respect to the past (i.e. quiescent; Log(SSFR) $<-10$, $b<<1$) and has units of inverse time. 
%SSFR=SFR/M SSFR is closely related to Scalo's birthrate parameter 'b', the ratio of the current SFR to the average SFR over all time.  \cite{kennicutt94} gives  b=SFR.\tau(1-R)/M_s Where Md and tau_d are the stellar mass and age and R is the fraction of stars returned for future star formation.
%Higher values of SSFR indicate that a larger fraction of stars were formed recently.  
Previous analyses have shown that star forming galaxies preferentially lie on a SSFR sequence with mass, such that the SSFR of low-mass galaxies is higher than for high-mass galaxies (e.g. \citealt{brinchmann04, schiminovich07, noeske07b}).  Galaxies in the Local Volume are also observed to lie on the SSFR--mass sequence \citep{bothwell09}. 

%SSFR is closely related to Scalo's stellar birthrate parameter \citep{kennicutt94}, $b=\frac{SFR}{Mass} \tau (1-R)$.  The age of the system, $\tau$, is generally assumed to be a Hubble time (13.7 Gyrs for our chosen cosmology) and the fraction of stars recycled for future star formation during the history of the system $R=0.4$ \citep{kennicutt94}.   
We calculate SFRs from the corrected H${\alpha}$ luminosity using the relationship given by \citet{kennicutt98}, i.e. $SFR (M_{\odot}\,{yr}^{-1})=L_{H\alpha} (W) / 1.27\times10^{34}$. %This conversion factor is calculated using a Salpeter IMF.
%, and stellar population models with solar metal abundance.  The calibrations also assume that the SFH is constant for at least the past $\sim10$ Myrs.

\begin{figure}
\begin{center}	
\resizebox{20pc}{!}{
\includegraphics{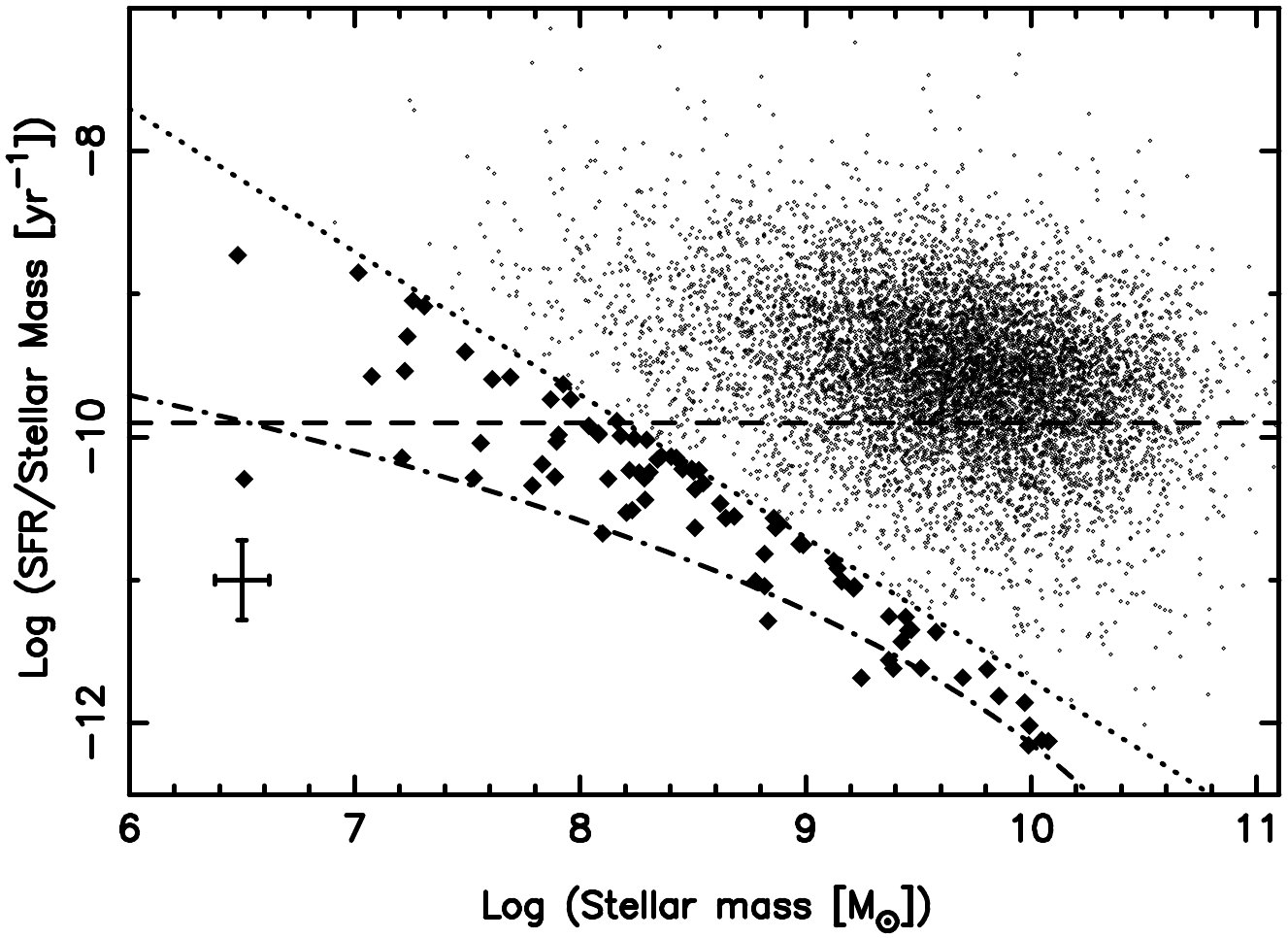}
}
\end{center}
\caption{Specific Star Formation Rate (SFR/Stellar Mass) as a function of Stellar Mass for the Main sample (star-forming galaxies with $0.002<z<0.13$).  
%The solid line indicates the SSFR-Mass relationship from \citet{schiminovich07}.  
The dashed line indicates the $b\sim1$ line for a constant star formation history over a Hubble time.  The dot-dashed line indicates the H$\alpha$ flux limit of our data. The dotted line indicates the upper H$\alpha$-luminosity limit of our sample. The lowest H$\alpha$-luminosity sample are indicated by the diamonds.  The median $1\sigma$ error on the stellar masses and SSFRs is given in the bottom left-hand corner.}
\label{ssfr_mass}
\end{figure}

The SSFR--stellar mass relationship is shown in Figure \ref{ssfr_mass}.  
%The relationship for UV-detected SDSS galaxies, shown by the solid line \citep{schiminovich07}.  
The SSFR consistent with a constant star formation history over a Hubble time (i.e. $b\sim1$) is  indicated by the dashed line.  The SSFR equivalent to our lower flux limit at the median redshift of our sample ($z\sim0.03$) is indicated by the dot-dashed line.  Our sample selection effectively selects galaxies at the limit in the SSFR - stellar mass relationship. 
%.  The low H$\alpha$-luminosity galaxies have the lowest SSFR at any stellar mass (due to their selection as low star-formation systems) 
However, the range of masses encompassed by our sample results in a range of SSFR, consistent with a range of star formation histories from quiescent to bursty, depending on their stellar mass.  Our flux limit prevents us from observing galaxies with lower SSFR, however, H$\alpha$ imaging surveys observe galaxies with SSFR below our flux limit at these masses (e.g.  \citealt{kennicutt94,lee07,bothwell09}).  The range of SFH present for these galaxies is therefore likely to be larger than the distribution we observe at the lowest stellar masses we probe.  It is also likely that there are more massive galaxies with similarly low star formation rates, as have been observed by GALEX (e.g. \citealt{jeong09,schiminovich10}).  This suggests that a simple relationship between SSFR and mass may be less warranted and may require more examination of the selection effects involved (e.g. \citealt{stringer10}).

\subsection{Environment}

Studies examining the distribution of H$\alpha$ luminosity (or EW(H$\alpha$)) with environment find that the relative proportions of galaxies with high H$\alpha$ luminosities depend on their environment, with more galaxies having high H$\alpha$ luminosities in less dense environments than in more dense environments \citep{lewis02,gomez03,doyle06,bamford08}.
%\cite{haines07} argue that the evolution of dwarf galaxies is heavily dependent on their environment, with passive dwarf galaxies only found in the densest of environments and star-forming dwarf galaxies only found outside of such environments.  
The depth and area of the GAMA survey mean that we can estimate whether the environment of the lowest  H$\alpha$ luminosity galaxies is responsible for the range of star formation histories we observe.   

We restrict the magnitude-limited Whole sample to the redshift range $0.02 < z < 0.13$, guaranteeing that all galaxies in the magnitude range $M_r<-19.5$ are volume limited.  We then calculate the distance for all galaxies in our sample to the nth nearest member of the volume-limited `Density Defining Population' (DDP; \citealt{croton05}), such that we do not over or underestimate the density for each galaxy as the redshift increases.  If the distance to the nearest neighbour is greater than that to the survey edge then the area used to calculate the density is that within a chord crossing the circle \citep{baldry06}.
%, unless the distance to the nearest neighbour is more than 5 times that to the survey edge, for which such an area is an inappropriate overestimate.  The 3 galaxies for which this is true are discarded from the density analysis.
%However, we have not made any completeness corrections so these densities provide lower limits on the final surface densities.

In order to get a measurement of density for as many of the low H$\alpha$ luminosity sample as possible, we examine the distance to the 1st nearest neighbour brighter than $M_r=-19.5$ mag within a velocity cylinder of $\pm1000$kms$^{-1}$ and obtain density measurements for 59 of our 90 low-H$\alpha$ luminosity galaxies. 
By reducing the nearest neighbour requirement from more usual numbers of 3-10 we get estimates of densities for a larger proportion of our sample.  While these measurements are less robust, they support the results we obtain for 3rd nearest neighbour measurements, which can only be made for 32 of our sample.  The 31 galaxies for which we cannot obtain density measurements are at redshifts $z<0.02$ and do not have neighbours within the volume limited sample, suggesting that they inhabit even lower-density regions.
%These values allow us to give a quantitative measurement of environment for the sample, however they are different enough from previous analyses (e.g.  \citealt{lewis02,croton05,baldry06}) to make direct comparison with these inappropriate.  However, 
We note that the conclusions we come to are not sensitive to the velocity cylinder length, the number of nearest neighbours or limiting magnitude used.  
%We also discard all galaxies whose nearest neighbour distance is less than that to the nearest survey edge.

\begin{figure}
\begin{center}	
%\resizebox{20pc}{!}{

%\includegraphics{../ssfr_dens}
%}
\resizebox{20pc}{!}{
\includegraphics{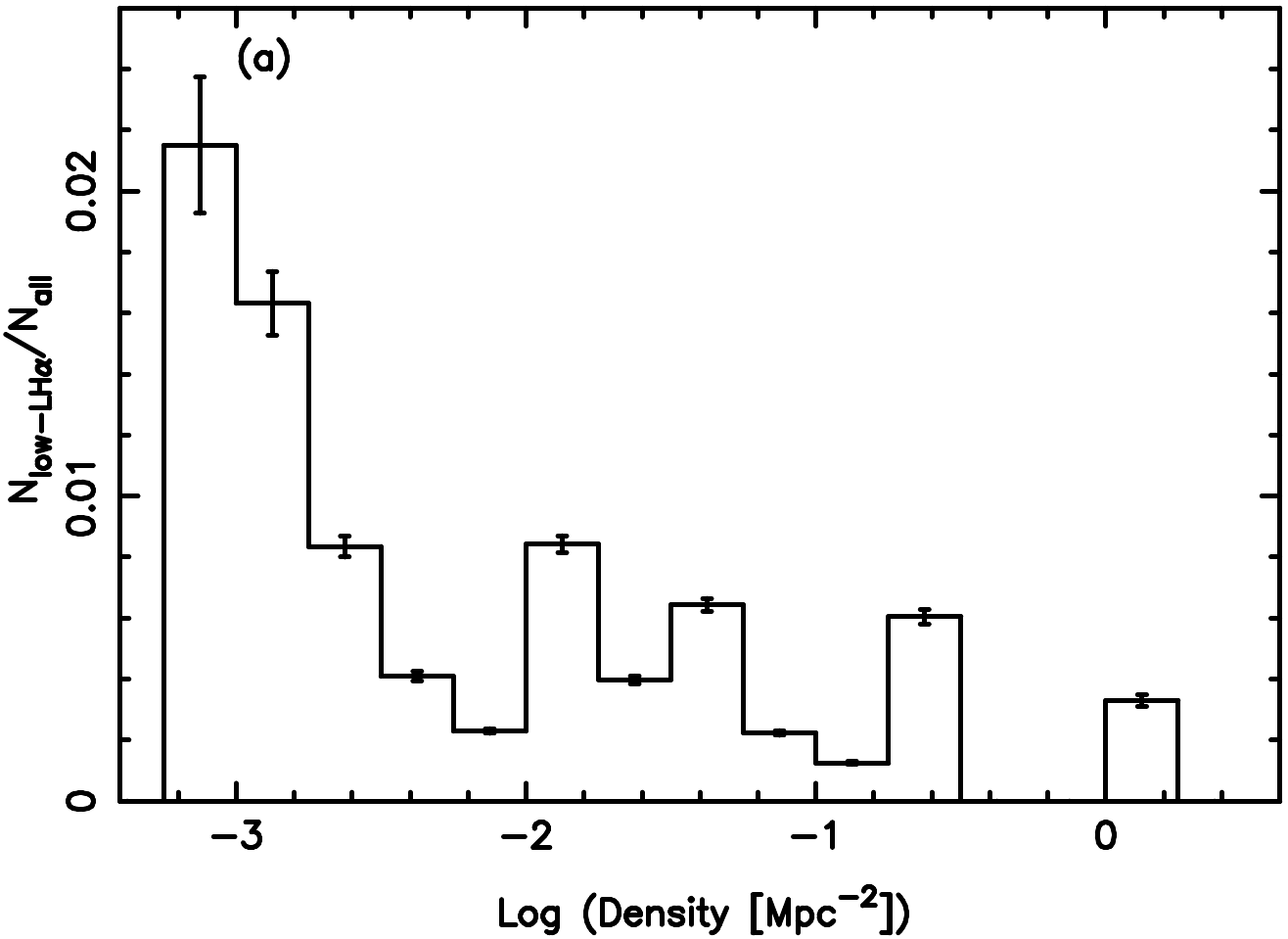}
}
\resizebox{20pc}{!}{

\includegraphics{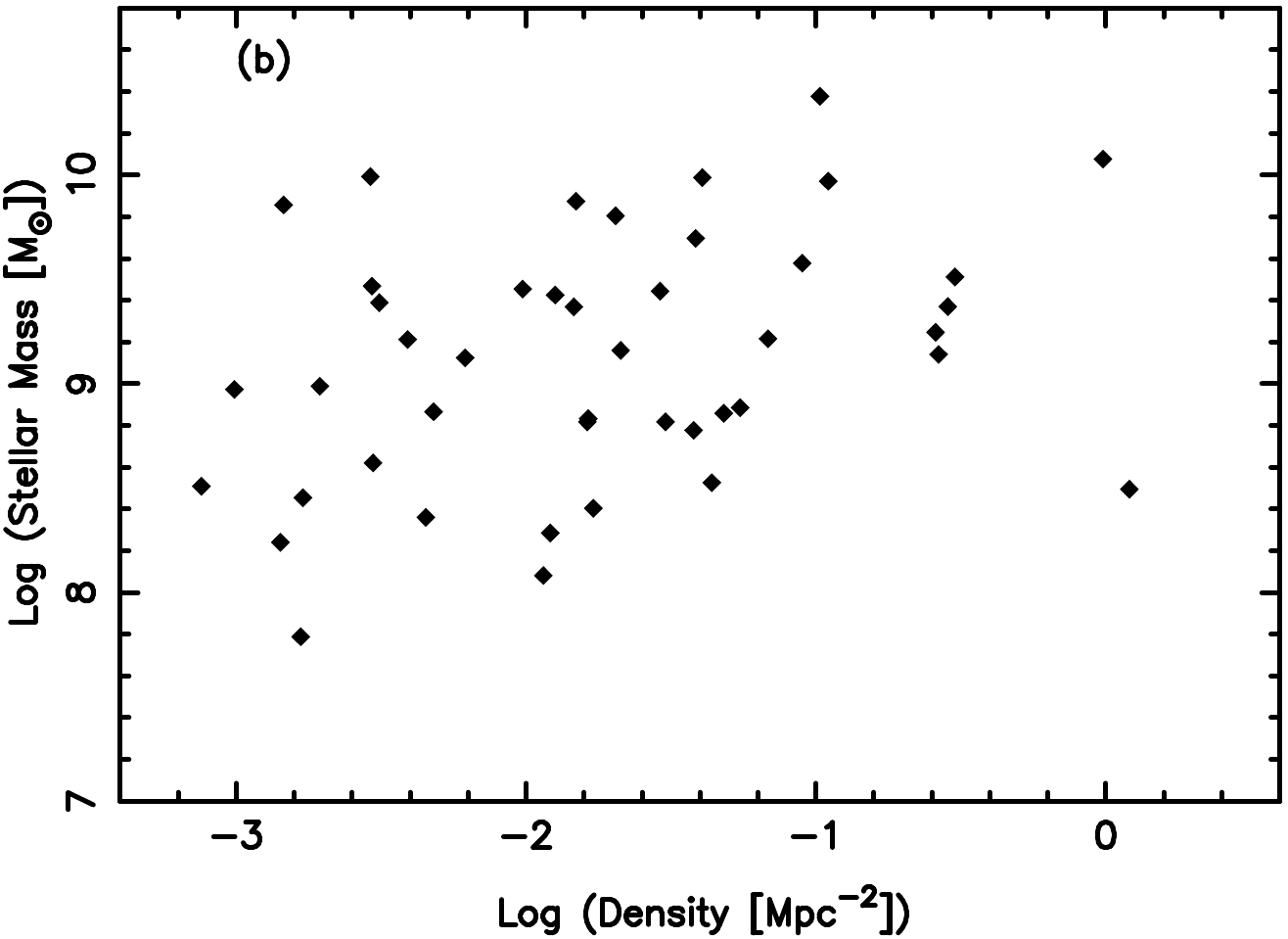}
}

\end{center}
\caption{
%Top: The distribution of the specific star formation rates of the main sample with surface density. The red diamonds indicate the low H$\alpha$ luminosity sample.  %The solid line indicates the cut made to select the lowest H$\alpha$ luminosity galaxies and the dashed line indicates the cut made to select an equal number of the highest H$\alpha$ luminosity galaxies.  
%Bottom: 
(a) Histogram of the fraction of low L$_{\rm{H}\alpha}$ galaxies as a function of all galaxies with the same range of stellar masses in the Whole sample (all galaxies with $0.002<z<0.13$) at each density}.  The low L$_{\rm{H}\alpha}$ sample are found at the lowest densities and are a higher fraction at the lowest densities.  The error bars indicate poisson errors on the number of galaxies in each density bin.  (b) Distribution of stellar mass with surface density for the lowest H$\alpha$ luminosity galaxies, emphasising that the lowest mass systems lie in the least dense environments.
\label{lha_dens}
\end{figure}

%Figure \ref{lha_dens}(Top) shows that the low-H$\alpha$ luminosity galaxies are found in a range of environments.  
In order to qualitatively compare the environments of the low-H$\alpha$ luminosity galaxies to those of other galaxies we 
%select a sample of passive (i.e. non-star-forming) 
compare to all galaxies with the same stellar mass range.   Our results are unchanged if we only compare to passive galaxies in the same mass range.  The lowest H$\alpha$ luminosity galaxies are found in the lowest density environments we can probe with a maximum density of only $1.5$ galaxy Mpc$^{-2}$ (median density $=0.016$ galaxy Mpc$^{-2}$).  In contrast, the Whole sample of galaxies are generally found in regions of much higher density, extending to $\sim1000$ galaxy Mpc$^{-2}$ (median density $=0.061$ galaxy Mpc$^{-2}$).  We show the fraction of the low-H$\alpha$ luminosity galaxies with respect to all galaxies as a function of their environmental density in Figure \ref{lha_dens}(a).   The fraction of the low-H$\alpha$ luminosity galaxies is significantly higher in the lowest density environment.  The population of galaxies that do not make it into our low-H$\alpha$ luminosity sample, but which also exist at these low densities, are mainly star-forming galaxies with higher H$\alpha$ luminosities than this sample.

%This suggests that the highest H$\alpha$ luminosity galaxies are found on the outskirts of clusters, consistent with previous observations \citep{lewis02,goto03,pimbblet06}.  
Figure \ref{lha_dens}(b) indicates that the environmental density of the low-H$\alpha$ luminosity galaxies is related to their stellar mass. For a given density the stellar mass is higher in higher density environments. There is only one low-mass galaxy in moderate density environments ($0.1>$ densities (galaxy Mpc$^{-2})>10$).  However, there are two higher mass ($M\sim10^{10}M_{\odot}$) low star formation galaxies in the lowest density environments.  The fact that there are no low-mass, low star formation galaxies in high density environments (density $>10$ galaxy Mpc$^{-2}$) 
%with those in the densest environments having the highest stellar masses.  {\bf Given the close relationship between SSFR and stellar mass seen in Figure~\ref{ssfr_mass} means that galaxies with the lowest mass have highest SSFR and reside in the lowest density environments.} This 
suggests that the only way such low-mass, star-forming, galaxies can survive over a Hubble time without being accreted, tidally disrupted or being starved of their gas supply is through residing in the lowest density environments.

These observations are consistent with measurements of galaxy clustering: The correlation length of the two-point correlation function is observed to increase from less clustered low-luminosity blue galaxies ($M_r>-19$) to more clustered high-luminosity blue galaxies ($M_r<-21$) and even more clustered red galaxies in the 2-degree Field Galaxy Redshift Survey \citep{norberg02} and SDSS (using photometric redshifts trained on GAMA observations; Christadoulou et al., in prep).  

%This relationship can be inferred from Figure \ref{mass_frac} which demonstrates the falling fraction of SF galaxies at higher stellar masses, where those stellar masses are generally found in higher density environments (e.g. \cite{croton05,baldry06}).

%COMMENT.

%\subsection{Metallicities}

%By examining the masses of galaxies in the next Halpha luminosity bin we can make a rough statement of the burstiness of these galaxies - with only X galaxies $<$ X mass with $X<H\alpha$Luminosity$<Y$.

\section{Discussion and Conclusions}
\label{sect:discussion}

We have investigated the properties of a sample of 90 of the faintest-H$\alpha$ luminosity galaxies from the $r-$band--selected GAMA survey.  We find these galaxies to generally be low-stellar mass, blue, in low-density environments with a range of star formation histories.  %They follow the same specific star-formation sequence as more massive galaxies.  They also follow a range of star formation histories.
%, however, at any given stellar mass the lowest H$\alpha$ luminosity galaxies have the more passive star formation history.
This is the most distant sample of such low-mass, low star formation systems currently known and, with the area of the GAMA survey, the best yet available for probing their environmental dependence.

%However, these galaxies are seen to be more slowly star forming with specific star formation rates consistent with their stellar mass forming over a Hubble time.  
%This suggests that starbursts are not the dominant mode of stellar mass production in dwarf galaxies.

%How do these galaxies compare to other galaxy populations?

%This latter mode, dominated by low-level and more-or-less continuous star formation, is similar to what we are observing in much lower mass systems.

We have shown that the low H$\alpha$ luminosity galaxies have similar properties to dwarf irregular galaxies in the Local Volume.  Local Group galaxies are sufficiently nearby that their star formation histories have been studied in detail through resolved stellar population analysis.  In this environment dwarf galaxies 
%($M_B<-16$ mag; \citealt{tammann94}) 
are the most numerous.  The Local Group dwarf galaxies appear to have complex, past star formation histories, such that whatever star formation is currently occurring, there is always evidence for large intermediate age populations and for long episodes of low-to-moderate star formation intensity separated by short passive phases (e.g. \citealt{smecker-hane96,skillman03,tolstoy09}).  %Gas accretion and outflows may play a role in creating the older stars and fluctuations (e.g. \citealt{koch06})

We find low-SFR, low-mass systems at higher redshifts than the Local Group/Volume samples yet with similar H$\alpha$ luminosities and wide ranges of star formation histories to the local samples.  
%These similar characteristics, here observed at higher redshifts suggest that the star-forming properties of these low-mass galaxies cannot evolve rapidly.  
The continuous distribution of SSFRs we observe suggests that the galaxies cover a range of phases of star formation, some passive, some steady and some bursting.  This is consistent with their being higher redshift analogues of local dwarf star-forming systems.  
%The fact that there is a continuous distribution in SSFR is evidence that the low-mass systems spend similar amounts of time in the `on' and `off' phases.  
%This provides further evidence for these low-mass galaxies having more or less constant levels of star formation over cosmic history. 
As introduced in \S \ref{sect:intro}, understanding the causes of this range in star formation histories is ongoing.  \cite{stinson07} simulate the collapse of isolated dwarf galaxies and find that star formation in low-mass galaxies can undergo a `breathing' mode where episodes of star formation trigger gas heating leading to galaxies with episodic star formation and significant intermediate age populations.  \cite{quillen08} show that this `breathing' requires strong, delayed feedback in order to reproduce the observationally estimated episode times.  Alternatively, we could have a steeper IMF as recently suggested for galaxies with lower SFRs (e.g. \citealt{weidner05,hoversten08,meurer09,lee09b,gunawardhana10}).

We find these galaxies to be in low-density environments.  That isolation has led to their remaining low mass and star forming, with the lowest mass galaxies generally found in the lowest density environments and those in denser environments tending to have higher stellar masses.  These observations lead to the prediction that at high-redshift, when there has been less time for interactions, we would expect to see a higher proportion of low-mass, star-forming systems in dense environments since they will not have had time to be accreted, tidally-stripped or their gas reservoirs rendered ineffective through external processes.

\section*{Acknowledgments}

We thank the anonymous referee for comments that improved the paper.  SB thanks Phil James, Baerbel Koribalski and Kim-Vy Tran for helpful comments.  GAMA is a joint European-Australasian project based around a spectroscopic campaign using the Anglo-Australian Telescope. The GAMA input catalogue is based on data taken from the Sloan Digital Sky Survey and the UKIRT Infrared Deep Sky Survey. Complementary imaging of the GAMA regions is being obtained by a number of independent survey programs including GALEX MIS, VST KIDS, VISTA VIKING, WISE, Herschel-ATLAS, GMRT and ASKAP providing UV to radio coverage. GAMA is funded by the STFC (UK), the ARC (Australia), the AAO, and the participating institutions. The GAMA website is http://www.gama-survey.org/ . 

%AMH acknowledges support provided by the Australian Research Council through a QEII Fellowship (DP0557850).  
%I thank an anonymous referee for very useful comments that improved the presentation of the paper.

%\bibliography{littlies}
%\begin{thebibliography}{99}
%\bibitem[\protect\citeauthoryear{Beichman et al.}{1985a}]{b2} Beichman
%C.A., Neugebauer G., Habing H.J., Clegg P.E., Chester T.J., 1985a,
%{\it IRAS\/} Point Source Catalog. Jet Propulsion Laboratory,
%Pasadena
%\end{thebibliography}

\bsp

\label{lastpage}

\end{document}